\documentclass[
aps,
pre,
superscriptaddress,
twocolumn,
showpacs,
showkeys,
]{revtex4-1}

\usepackage{graphicx}
\usepackage{dcolumn}
\usepackage{bm}
\usepackage{subfigure}
\usepackage{amsthm}\newtheorem{definition}{Definition}
\usepackage{CJK}
\usepackage{tabularx}
\usepackage{multirow}

\begin{document}
\preprint{Physical Review Letters}

\title{Identification of influential nodes in network of networks}


\author{Meizhu Li}
\affiliation{School of Computer and Information Science, Southwest University, Chongqing 400715, China}
\author{Qi Zhang}
\affiliation{School of Computer and Information Science, Southwest University, Chongqing 400715, China}
\author{Qi Liu}
\affiliation{Center for Quantitative Sciences, Vanderbilt University School of Medicine, Nashville, TN 37232, USA}
\affiliation{Department of Biomedical Informatics, Vanderbilt University School of Medicine, Nashville, TN 37232, USA}

\author{Yong Deng}
\email{ydeng@swu.edu.cn; prof.deng@hotmail.com}
\affiliation{School of Computer and Information Science, Southwest University, Chongqing 400715, China}%
\affiliation{School of Automation, Northwestern Polytechnical University, Xian, Shaanxi 710072, China}
\affiliation{School of Engineering, Vanderbilt University, Nashville, TN, 37235, USA}%



\date{\today}

\begin{abstract}
    The network of networks(NON) research is focused on studying the properties of $n$ interdependent networks which is ubiquitous in the real world. Identifying the influential nodes in the network of networks is theoretical and practical significance. However, it is hard to describe the structure property of the NON based on traditional methods. In this paper, a new method is proposed to identify the influential nodes in the network of networks base on the evidence theory. The proposed method can fuse different kinds of relationship between the network components to constructed a comprehensive similarity network. The nodes which have a big value of similarity are the influential nodes in the NON. The experiment results illustrate that the proposed method is reasonable and significant.
\end{abstract}



\maketitle

\section{Introduction}
Complex networks describe a wide range of systems in nature and society, it has been widely used in many fields \cite{newman2003structure,newman2006structure,watts1998collective}. In the real world, a large amount of systems can be described as the complex networks, such as the internet, airline routes, electric power grids and the protein interaction networks. The function of all these networks relies on the connectivity between the network components. However in the real world, numbers of networks have the property that the nodes of the network have different kinds of relationship based on different principles, such as the protein interaction networks and the cancer gene expression network \cite{nicosia2013growing,wang2014similarity,kenett2014network}. This kind of networks is called the network of networks (NON) \cite{gao2011robustness}.

Compared to the traditional complex networks with single relationship, the network of networks is more difficult to illuminate the structure property of it \cite{boccaletti2014structure,battiston2014structural,Zhang2015707}. In the NON, identification of the influential nodes is theoretical and practical significance. In this paper, a new method is proposed to identify the influential nodes in the network of networks based on the evidence theory.

Dempster-Shafer theory of evidence \cite{dempster1967upper,shafer1976mathematical}, is used to deal with uncertain information and has been widely used in many fields \cite{bloch1996some,cuzzolin2008geometric,denoeux2011maximum,Deng2011,li2014multiscale}. Here the combination rules of evidence theory are used to fuse the influence of each node in different single networks. The nodes which have a big value of similarity are the influential nodes in the NON. One of the advantages of the proposed method is that the more the type of the interrelation between networks components is, the more accurate the results are.

The rest of this paper is organised as follows. Section \ref{Rreparatorywork} introduces some preliminaries of this work. In section \ref{new}, a new method is proposed to identify the influential nodes in the network of network. The application based on the cancer gene expression networks is illustrated in section \ref{application}. Conclusion is given in Section \ref{conclusion}.

\section{Preliminaries}
\label{Rreparatorywork}
\subsection{The network of networks}
\label{network of networks}
The network of networks(NON), sometimes called multilayer networks or multiplex, has attracted more and more attention. Due to the fast growth of this field, there are many definitions of different types of NON, such as interdependent networks, interconnected networks, multilayered networks, multiplex networks and so on. There exist many datasets that can be represented as NON, such as flight networks, reliway networks and road networks, network of biological networks including gene regulation networks, metabolic network and protein-protein interacting network.

In this paper, we focus on the network which have the same nodes and different kinds of relationship between the components in the networks.

\subsection{The evidence theory}
\label{evidence theory}
 Dempster-Shafer theory \cite{dempster1967upper,shafer1976mathematical} is often regarded as an extension of the bayesian theory. For completeness of the explanation, a few basic concepts are introduced as follows.

\begin{definition}
Let $\Omega$ be a set of mutually exclusive and collectively
exhaustive, indicted by

\begin{footnotesize}
\begin{equation}
\Omega  = \{ E_1 ,E_2 , \cdots ,E_i , \cdots ,E_N \}
\end{equation}
\end{footnotesize}

The set $\Omega$ is called frame of discernment. The power set of
$\Omega$ is indicated by $2^\Omega$, where

\begin{footnotesize}
\begin{equation}
2^\Omega   = \{ \emptyset ,\{ E_1 \} , \cdots ,\{ E_N \} ,\{ E_1
,E_2 \} , \cdots ,\Omega \}
\end{equation}
\end{footnotesize}

If $A \in 2^\Omega$, $A$ is called a proposition.
\end{definition}

\begin{definition}
For a frame of discernment $\Omega$,  a mass function is a mapping
$m$ from  $2^\Omega$ to $[0,1]$, formally defined by:

\begin{footnotesize}
\begin{equation}
m: \quad 2^\Omega \to [0,1]
\end{equation}
\end{footnotesize}

which satisfies the following condition:

\begin{footnotesize}
\begin{eqnarray}
m(\emptyset ) = 0 \quad and \quad \sum\limits_{A \in 2^\Omega }
{m(A) = 1}
\end{eqnarray}
\end{footnotesize}

\end{definition}

In Dempster-Shafer theory, a mass function is also called a basic
probability assignment (BPA).

Consider two pieces of evidence indicated by two BPAs $m_1$ and
$m_2$ on the frame of discernment $\Omega$, Dempster's rule of
combination is used to combine them. This rule assumes that these
BPAs are independent.

\begin{definition}
Dempster's rule of combination, also called orthogonal sum,
denoted by $m = m_1 \oplus m_2$, is defined as follows

\begin{footnotesize}
\begin{equation}
m(A) = \left\{ {\begin{array}{*{20}l}
   {\frac{1}{{1 - K}}\sum\limits_{B \cap C = A} {m_1 (B)m_2 (C)} \;,} & {A \ne \emptyset ;}  \\
   {0\;,} & {A = \emptyset }.  \\
\end{array}} \right.
\end{equation}
\end{footnotesize}

with

\begin{footnotesize}
\begin{equation} \label{q_8}
K = \sum\limits_{B \cap C = \emptyset } {m_1 (B)m_2 (C)}
\end{equation}
\end{footnotesize}

\end{definition}

where $B$ and $C$ are also elements of $2^\Omega$, and K is a constant to show the conflict between
the two BPAs.

Note that the Dempster's rule of combination is only applicable to
such two BPAs which satisfy the condition $K < 1$.

\section{New methods to identify the influential nodes in the network of networks}
\label{new}
To identify the influential nodes in the complex networks is one of the important directions in network science. In the single network, many methods have been proposed to identify the influential nodes, such as the degree centrality, the betweenness centrality, the local structure entropy and so on. However, in the network of networks, more than two networks depend on each other, the structure of it becomes more complex. Based on the evidence theory, a new method is proposed to identify the influential nodes in NON by making a combination among the networks divided from the NON.

\begin{figure}[htbp]
  \centering
  \includegraphics[scale=1]{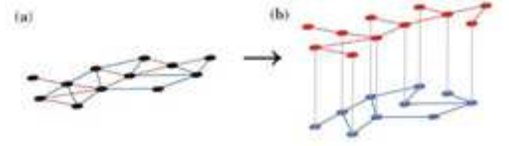}\\
  \caption{Division of NON.}\label{Show_NON}
\end{figure}

 As shown in the subfigure (a) of Fig.\ref{Show_NON}, our research is focus on the NON which has the same nodes but different edges. In the proposed method, one NON can be divided into numbers of single networks based on different principles, which is shown in the subfigre (b) of Fig.\ref{Show_NON}. Based on the single networks divided from NON, a series of similarity networks can be established. According to the similarity networks, a comprehensive networks can be constructed by fusing the similarity networks based on the combination rules of evidence theory. Compared to other nodes in the comprehensive networks, the nodes have a large value of similarity are the influential nodes in the NON.

 To introduce the proposed method in details, four steps are essential in the identification research, which is shown as follows.

 \begin{description}
   \item[Step 1]  Based on the different significance of the edges , divide the NON into numbers of single networks.

   \item[Step 2] According to the distance between each node, establish the distance matrix ${D}$ of the single networks. Each single network has a distance matrix ${D}$ to describe the similarity between each node. The details of the distance matrix ${D}$ are shown as follows:

       \[D = \left( {\begin{array}{*{20}{c}}
  {{d_{11}}}& \ldots &{{d_{1n}}} \\
   \vdots & \ddots & \vdots  \\
  {{d_{n1}}}& \cdots &{{d_{nn}}}
\end{array}} \right)\]

    \[{d_{max }} = \max ({d_{ij}}),(1 \leq i \leq n,1 \leq j \leq n,)\]

    Where ${d_{ij}}$ is the shortest distance between node $i$ and node $j$.

    Based on the distance matrix, the similarity network $SN$ can be defined as follows:

    \[SN = \left( {\begin{array}{*{20}{c}}
  {{S_{11}}}& \ldots &{{S_{1n}}} \\
   \vdots & \ddots & \vdots  \\
  {{S_{n1}}}& \cdots &{{S_{nn}}}
\end{array}} \right)\]

\[{S_{ij}} = 1 - \frac{{{d_{ij}}}}{{{d_{\max }}}}\]

   \item[Step 3] According to similarity network, the basic probability assignment(BPA) can be constructed, which is an essential concept in the evidence theory. Each element in the similarity network has a corresponding BPA, which is defined as follows.
       \begin{definition}\label{d4}
       Given an $N \times N$ similarity network $SN$, the frame of discernment of the network is $\{Y,N\}$, where $Y$ represents similarity and $N$ represents dissimilarity. The BPA of element $SN_{ij}$ is:
\begin{footnotesize}
       \begin{equation}
       {m_{ij}(Y)} = \frac{{\left| {S{N_{ij}} - min(SN)} \right|}}{{SUMM}}
       \end{equation}

       \begin{equation}
       {m_{ij}(N)} = \frac{{\left| {S{N_{ij}} - \max (SN)} \right|}}{{SUMM}}
       \end{equation}

       \begin{equation}
       {m_{ij}}(Y,N) = \frac{{\left| {S{N_{ij}} - (\max (SN) + \min (SN))/2} \right|}}{{SUMM}}
       \end{equation}

       \begin{figure*}
       \begin{equation}
       SUMM = \left| {S{N_{ij}} - \max (SN)} \right| + \left| {S{N_{ij}} - \min (SN)} \right| + \left| {S{N_{ij}} - (\max (SN) + \min (SN))/2} \right|
       \end{equation}
       \end{figure*}
\end{footnotesize}
       Where $max(SN)$ represents the maximum element in the similarity network, except the diagonal elements. $min(SN)$ represents the minimum element in the similarity network.
       \end{definition}

   Based on the combination rules of the evidence theory, fuse the BPA of corresponding elements in similarity network into a comprehensive similarity network.

   \item[Step 4]  The nodes which have a big value of similarity with other nodes in the fused similarity network are the influential nodes in the NON.

 \end{description}

\begin{figure}[htbp]
  \centering
  \includegraphics[scale=0.8]{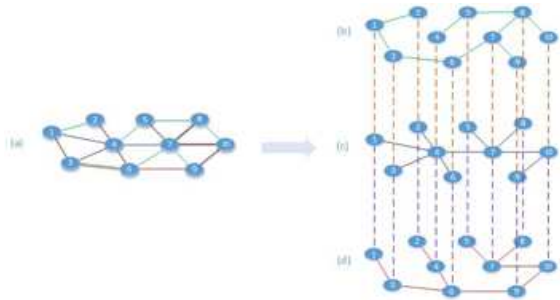}\\
  \caption{Example of NON and the corresponding divied single networks.}\label{Show_E_NON}
\end{figure}

Here a network of networks, which is shown in Fig.\ref{Show_E_NON}, is constructed to show the details of the combination process.

The NON in the Fig.\ref{Show_E_NON} can be divided into three single networks. The details of the network (b), network (c) and network (d) are shown as follows.

\begin{footnotesize}
\[\begin{array}{l}
{V_b}={V_c}={V_d} = \{ 1,2,3,4,5,6,7,8,9,10\} \\
{E_b} = \{ \{ 1,2\} ,\{ 1,3\} ,\{ 3,6\} ,\{ 4,5\} ,\{ 5,8\} ,\{ 6,7\} ,\{ 7,8\} ,\{ 7,9\} ,\{ 8,10\} \} \\
{E_c} = \{ \{ 1,4\} ,\{ 2,4\} ,\{ 3,4\} ,\{ 4,6\} ,\{ 4,7\} ,\{ 5,7\} ,\{ 7,8\} ,\{ 7,10\} ,\{ 9,10\} \} \\
{E_d} = \{ \{ 1,3\} ,\{ 2,4\} ,\{ 3,6\} ,\{ 4,6\} ,\{ 5,7\} ,\{ 6,9\} ,\{ 7,8\} ,\{ 7,10\} ,\{ 9,10\} \}
\end{array}\]
\end{footnotesize}

The similarity matrix of the three single networks is shown as follows.

The similarity matrix of the single network (b):

\begin{footnotesize}
\[\left[
\begin{array}{cccccccccc}
 1.00&0.86&0.86&0.15&0.29&0.72&0.58&0.43&0.43&0.29 \\
 0.86&1.00&0.72&0.01&0.15&0.58&0.43&0.29&0.29&0.15  \\
 0.86&0.72&1.00&0.29&0.43&0.86&0.72&0.58&0.58&0.43 \\
 0.15&0.01&0.29&1.00&0.86&0.43&0.28&0.72&0.43&0.58 \\
 0.29&0.15&0.43&0.86&1.00&0.58&0.72&0.86&0.58&0.72 \\
 0.72&0.58&0.86&0.43&0.58&1.00&0.86&0.72&0.72&0.58 \\
 0.58&0.43&0.72&0.58&0.72&0.86&1.00&0.86&0.86&0.72 \\
 0.43&0.29&0.58&0.72&0.86&0.72&0.86&1.00&0.72&0.86 \\
 0.43&0.29&0.58&0.43&0.58&0.72&0.86&0.72&1.00&0.58 \\
 0.29&0.15&0.43&0.58&0.72&0.58&0.72&0.86&0.58&1.00
\end{array}
\right]\]
\end{footnotesize}

\begin{footnotesize}
The similarity matrix of the single network (c):
\[\left[
\begin{array}{cccccccccc}
 1.00&0.51&0.51&0.75&0.26&0.51&0.51&0.26&0.01&0.26 \\
 0.51&1.00&0.51&0.75&0.26&0.51&0.51&0.26&0.01&0.26  \\
 0.51&0.51&1.00&0.75&0.26&0.51&0.51&0.26&0.01&0.26 \\
 0.75&0.75&0.75&1.00&0.51&0.75&0.75&0.51&0.26&0.51 \\
 0.26&0.26&0.26&0.51&1.00&0.26&0.75&0.51&0.26&0.51 \\
 0.51&0.51&0.51&0.76&0.26&1.00&0.51&0.26&0.01&0.26 \\
 0.51&0.51&0.51&0.76&0.76&0.51&1.00&0.76&0.51&0.75 \\
 0.26&0.26&0.26&0.51&0.51&0.26&0.76&1.00&0.26&0.51 \\
 0.01&0.01&0.01&0.26&0.26&0.01&0.51&0.26&1.00&0.76 \\
 0.26&0.26&0.26&0.51&0.51&0.26&0.76&0.51&0.76&1.00
\end{array}
\right]\]
\end{footnotesize}

\begin{footnotesize}
The similarity matrix of the single network (d):
\[\left[
\begin{array}{cccccccccc}
 1.00&0.34&0.84&0.51&0.01&0.67&0.17&0.01&0.51&0.34 \\
 0.34&1.00&0.51&0.84&0.01&0.67&0.17&0.01&0.51&0.34  \\
 0.84&0.51&1.00&0.67&0.17&0.84&0.34&0.17&0.67&0.51 \\
 0.51&0.84&0.67&1.00&0.17&0.84&0.34&0.17&0.67&0.51 \\
 0.01&0.01&0.17&0.17&1.00&0.34&0.84&0.67&0.51&0.67 \\
 0.67&0.67&0.84&0.84&0.34&1.00&0.51&0.34&0.84&0.67 \\
 0.18&0.18&0.34&0.34&0.84&0.51&1.00&0.84&0.67&0.84 \\
 0.01&0.01&0.17&0.17&0.67&0.34&0.84&1.00&0.51&0.67 \\
 0.51&0.51&0.67&0.67&0.51&0.84&0.67&0.51&1.00&0.84 \\
 0.34&0.34&0.51&0.51&0.67&0.67&0.84&0.67&0.84&1.00
\end{array}
\right]\]
\end{footnotesize}

According to Definition \ref{d4}, the BPA of each element in the similarity networks can be constructed. Then using the combination rules of evidence theory, fuse the corresponding BPA of each element in the similarity networks. Here an example is shown to fuse the element $SN_{34}$ of network (b), network (c) and network (c).

Based on the three single networks above, the values of the element $SN_{34}$ in the networks can be shown as follows.

$SN_{34}^b = 0.29$, $SN_{34}^c = 0.75$, $SN_{34}^d = 0.67$.

According to Definition \ref{d4}, the BPA of elements $SN_{34}$ in the networks can be constructed.

\begin{small}

$m_{34}^b(Y) = 0.2814$, $m_{34}^b(N) = 0.5729$, $m_{34}^b(Y,N) = 0.1457$.

$m_{34}^c(Y) = 0.6637$, $m_{34}^c(N) = 0.0090$, $m_{34}^c(Y,N) = 0.3273$.

$m_{34}^d(Y) = 0.6140$, $m_{34}^d(N) = 0.1581$, $m_{34}^d(Y,N) = 0.2279$.

\end{small}

After combination based on evidence theory, the fusion result of element $SN_{34}$ is:

\begin{small}

$m_{34}(Y) = 0.7874$, $m_{34}(N) = 0.1878$, $m_{34}(Y,N) = 0.0248$.

\end{small}

So the value of similarity between node 3 and node 4 in the network and networks is $0.7874$. The order of the influential nodes in the example networks is shown in Table \ref{tab:example}:

\begin{table}[htbp]
  \centering
  \caption{The influential nodes in the example network of networks}
    \begin{tabular}{ccccccccccc}
    \hline
    Order number   & 1     & 2     & 3     & 4     & 5     & 6     & 7     & 8     & 9     & 10 \\
    \hline
    Node number   & 7   & 6   & 4   & 10   & 3    & 9    & 8   & 5    & 1   & 2 \\
    \hline
    \end{tabular}%
  \label{tab:example}%
\end{table}%

\section{Application of the new method in the cancer gene expression networks }
\label{application}
In order to illuminate the usefulness of the proposed method, four cancer gene expression networks, the glioblastoma multiforme (GBM), the breast invasive carcinoma (BIC), the kidney renal clear cell carcinoma (KRCCC) and the lung squamous cell carcinoma (LSCC), have been applied as cases. Each cancer gene expression network has three kinds of expression networks, the DNA methylation network, the mRNA expression network and the miRNA expression network.

The nodes in the networks represent the patients. The relationship in these expression networks are the similarity between each patient. Based on the definition of the network of networks , the four cancer gene expression networks can treated as four NON. In order to find the influential patients in the NON, a comprehensive similarity expression network is constructed by the combination rules of evidence theory. Each NON in this case has three single networks, the DNA methylation network, the mRNA expression network and the miRNA expression network.

\begin{figure}[htbp]
  \centering
  \subfigure[GBM]{
    \label{Local_detail:subfig:a} 
    \centering
    \includegraphics[scale=2.15]{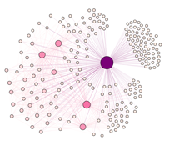}}
  \hspace{0.5cm}
    \subfigure[BIC]{
    \label{Local_detail:subfig:b} 
    \centering
    \includegraphics[scale=2.15]{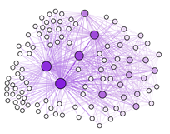}}
  \hspace{0.5cm}
    \subfigure[KRCCC]{
    \label{Local_detail:subfig:c} 
    \centering
    \includegraphics[scale=2.15]{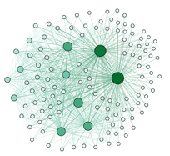}}
  \hspace{0.5cm}
      \subfigure[LSCC]{
    \label{Local_detail:subfig:d} 
    \centering
    \includegraphics[scale=2.15]{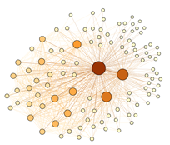}}
  \caption{Influential nodes of four NON}\label{Local_detail}
\end{figure}

The experiment results are shown in the Table \ref{tab:addlabel} and Fig. \ref{Local_detail}. In the Table \ref{tab:addlabel}, the number of most influential nodes in the four NON are shown. For example, in the network GBM, the first influential node is the node 116. In the Fig. \ref{Local_detail}, the nodes with bigger size and deeper color are the more influential nodes. The results illustrate that the proposed method is reasonable and significant.

\begin{table}[htbp]
  \centering
  \caption{The influential nodes in the four cancer gene expression networks}
    \begin{tabular}{ccccccccccc}
    \hline
    Order number   & 1     & 2     & 3     & 4     & 5     & 6     & 7     & 8     & 9     & 10 \\
    \hline
    GBM   & 116   & 60    & 190   & 179   & 50    & 42    & 139   & 72    & 194   & 209 \\
    BIC   & 106   & 34    & 71    & 49    & 68    & 7     & 55    & 76    & 100   & 51 \\
    KRCCC & 104   & 15    & 113   & 117   & 110   & 50    & 41    & 96    & 77    & 87 \\
    LSCC  & 37    & 90    & 91    & 51    & 93    & 33    & 88    & 28    & 92    & 38 \\
    \hline
    \end{tabular}%
  \label{tab:addlabel}%
\end{table}%

\section{Conclusion}
\label{conclusion}
Many real systems in the real world can be treated as the network of networks. Identifying the influential nodes in the network of networks is theoretical and practical significance. Dempster-Shafer theory of evidence is used to deal with uncertain information and has been widely used in many fields. In this paper the combination rules of evidence theory are used to fuse the influence of each node in different single networks. The proposed method can fuse different kinds of relationship between the network components to constructed a comprehensive similarity network. The nodes which have a big value of similarity are the influential nodes in the NON. One of the advantages of the proposed method is that the more the type of the interrelation between networks components is, the more accurate the results are. The experiment results illustrate that the proposed method is reasonable and significant.

\section{Acknowledgment}
The work is partially supported by National High Technology Research and Development Program of China (863 Program) (Grant No. 2013AA013801), National Natural Science Foundation of China (Grant No. 61174022), Specialized Research Fund for the Doctoral Program of Higher Education (Grant No. 20131102130002), R\&D Program of China (2012BAH07B01), the open funding project of State Key Laboratory of Virtual Reality Technology and Systems, Beihang University (Grant No.BUAA-VR-14KF-02). Fundamental Research Funds for the Central Universities No. XDJK2015D009. Chongqing Graduate Student Research Innovation Project (Grant No. CYS14062).
\bibliography{NONreference}

\end{document}